\documentclass[dvips]{article}

\usepackage{icrc2011}
\usepackage{float}
\restylefloat{table}

\title{Abundances of Suprathermal Heavy Ions in CIRs on STEREO during the Minimum of Solar Cycle 23}

\newcommand{\etal}{\MakeLowercase{\textit{et al. }}} 
\shorttitle{R. Bu\v{c}\'{\i}k \etal CIR abundances}

\authors{R. Bu\v{c}\'{\i}k$^{1}$, U. Mall$^{1}$, A. Korth$^{1}$, G. M. Mason$^{2}$}
\afiliations{$^1$Max-Planck-Institut f\"ur Sonnensystemforschung, Katlenburg-Lindau, Germany\\ $^2$Applied Physics Laboratory, Johns Hopkins University, Laurel, MD, USA}
\email{bucik@mps.mpg.de}

\abstract{We examine the composition of the $\sim$ 0.1-1 MeV/n interplanetary heavy ions from H to Fe in corotating interaction regions (CIRs) measured by the SIT (Suprathermal Ion Telescope) instrument. We use observations taken on board the two STEREO spacecraft during the unusually long minimum of Solar Cycle 23 from January 2007 through December 2010. During this period instruments on STEREO observed more than 50 CIR events making it possible to investigate CIR ion abundances during solar minimum conditions with unprecedentedly high statistics. The observations reveal annual variations of relative ion abundances in the CIRs during the 2007-2008 period. In 2010 the elemental composition in CIRs were influenced by solar energetic particle events.}

\begin{document}
\maketitle

\section{Introduction}

Energetic ions accelerated in corotating interaction regions (CIRs) have elemental abundances very close to the fast solar wind composition, except for the overabundance of $^{4}$He, Ne and C \cite{lab1}. The overabundance of $^{3}$He has been recently reported by Mason \etal \cite{lab2}, suggesting that the remnant impulsive flare ions are accelerated in CIRs. The singly ionized interstellar He, Ne and inner source C pick-up ions provide another population which is accelerated in CIRs \cite{lab3,lab4}. Although many compositional features of the suprathermal heavy ions in CIRs are known the relative contribution from different sources is not well understood.

In this paper we report the elemental abundances of the suprathermal H, He, O, NeS and Fe in CIRs and discuss event-to-event variations of the abundance ratios over the long solar minimum period between January 2007 and December 2010. 

\section{Observations}

The measurements presented here were made with the Suprathermal Ion Telescope (SIT) instruments \cite{lab5} onboard the STEREO-A and -B spacecraft. The SIT instrument is a time-of-flight mass spectrometer which measures ions from H to Fe in the energy range from 20 keV/n to several MeV/n. 

 \begin{figure*}[!t]
  \vspace{-6 mm}
  \centering
  \includegraphics[width=13. cm]{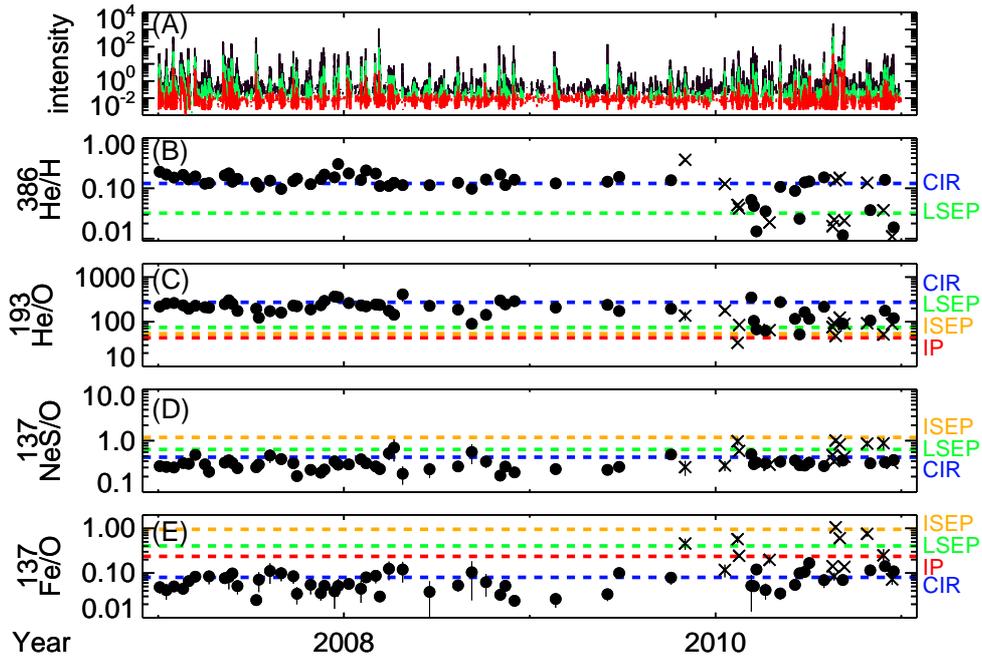}
  \caption{Panel (A): SIT/STEREO-A 1-hr averaged He intensity \#/(cm$^{2}$\,s\,sr\,MeV/n) for 189, 384, and 787 keV/n. Panels (B - E): Event-averaged 386 keV/n He/H, 193 keV/n He/O, 137 keV/n NeS/O and 137 keV/n Fe/O elemental ratios. CIR events are marked by {\it filled circles}, SEP events by {\it crosses}. {\it Dashed lines} show abundances measured in various particle populations present in the heliosphere: CIR events ({\it blue}), LSEP events ({\it green}), ISEP events ({\it yellow}) and IP shocks events ({\it red}).}
  \label{fig01}
 \end{figure*}

\begin{table*}[!t]
\begin{center}
\begin{tabular}{ccccccc}
\hline
      & CIR$^{a}$         &  CIR$^{b}$       &  LSEP$^{c}$       &     LSEP$^{d}$    & ISEP$^{e}$        &  IP shock$^{f}$\\
      &  (150 keV/n)      & (385 keV/n)      &  (358 keV/n)      &  ($>$300 keV/n)   & (358 keV/n)       &  (750 keV/n) \\
\hline
He/H  & [0.125$\pm$0.065] &   --             &   --              & [0.032$\pm$0.003] &   --              &    --          \\
He/O  & 113$\pm$20        & [273$\pm$72]     & [75.0$\pm$23.6]   & 52$\pm$4          &  [54$\pm$14]      & [44.4$\pm$14.4] \\
NeS/O &	0.48$\pm$0.13	  & [0.477$\pm$0.017]& [0.675$\pm$0.020] &	--           &	[1.158$\pm$0.022]& 0.678$\pm$0.014 \\
Fe/O  &	[0.08$\pm$0.03]   & 0.088$\pm$0.007  & [0.404$\pm$0.047] & 0.24$\pm$0.03     &	[0.95$\pm$0.005] & [0.236$\pm$0.01]\\
\hline
\end{tabular}
\begin{tabular}{l}
\hspace{-18mm}
Note: The ratios in square brackets correspond to the dashed lines in Figure \ref{fig01}.\\
\hspace{-18mm}
$^{a}$Average of 17 CIR events during solar minimum between December 1992 and July 1995 \cite{lab6}.\\
\hspace{-18mm}
$^{b}$Average of 41 CIR events during Solar cycle 23 betwen November 1, 1997 and June 1, 2007 \cite{lab2}.\\
\hspace{-18mm}
$^{c}$Average of 64 LSEP events between November 1997 and January 2005 \cite{lab7}.\\
\hspace{-18mm}
$^{d}$Average of 10 LSEP events between late 1977 and early 1981 \cite{lab8}.\\
\hspace{-18mm}
$^{e}$Average of 20 ISEP events between September 1997 and April 2003 \cite{lab9}.\\
\hspace{-18mm}
$^{f}$Avearge of 72 IP shocks between October 1997 and September 2002 \cite{lab10}.

\end{tabular}

\caption{Heavy ion abundances.}\label{table1}
\end{center}

\end{table*}

Figure \ref{fig01} provides an overview of the data over the period from January 2007 to December 2010. During the investigated period the monthly mean sunspot number taken from the NOAA never exceed number 15, indicating a low solar activity level. Panel (A) shows suprathermal 1-hr averaged He ion intensity \#/(cm$^{2}$\,s\,sr\,MeV/n) for 189, 384, and 787 keV/n measured by SIT-A. Panels (B - E) show event-integrated abundance ratios He/H, He/O, NeS/O and Fe/O. Shown are abundances where 1-hr averaged 189 keV/n He ion intensities exceed 5 particles/(cm$^{2}$\,s\,sr\,MeV/n). We use SIT pulse height analysis data to determine the abundance ratios. The horizontal dashed lines present average abundances for CIR \cite{lab2,lab6}, large solar energetic particle (LSEP) \cite{lab7,lab8}, impulsive SEP \cite{lab9} and interplanetary shock (IP) \cite{lab10} events listed in Table \ref{table1}. The filled circles in Figure \ref{fig01} indicate CIR events. We identify CIR events using the list of the CIRs compiled by the STEREO magnetometer team at the University of California Los Angeles. The CIR events in the period January 2007-September 2009 were examined in \cite{lab11}. The events marked by crosses show sharp rise in the intensity of the relativistic electrons. The list of the STEREO electron events has been compiled by the SEPT instrument team at Universit\"{a}t Kiel. Taken together with the elemental abundances the intensity increases marked by crosses are likely related to SEP events.

Figure \ref{fig01} shows that the CIR event He/H ratios during the period 2007-2009 are consistent with the average He/H ratio, observed in the previous solar minimum \cite{lab6}. Large event-to-event variations of the He/H in the CIRs occurred in 2010 when the SEP event activity considerably increased. In 2007-2009 the CIR event He/O ratios were close to the average He/O ratio obtained in the earlier surveys \cite{lab2,lab6}. The CIR event He/O ratio showed also large spread in 2010. In contrast to the He/H and He/O there is no observed increase in scatter of the CIR Fe/O and NeS/O ratios in 2010. Notice in Figure \ref{fig01} that the CIR NeS/O ratio stays relatively constant in 2010.

An interesting feature seen in Figure \ref{fig01} is the shape of the temporal variation of the CIR event Fe/O ratios between January 2007 and the beginning of 2009. The Fe/O ratios show local minima near the beginning and end of 2007 and near the end of 2008. Although there is some scatter in the data points the local maxima in the Fe/O have a tendency to occur in the middle of 2007 and 2008. Another interesting feature is that during the Fe/O minimum at the end of 2007 the He/O and He/H ratios show enhancements. The behavior seen in the ratios He/H, He/O and Fe/O is not apparent in the temporal profile of the NeS/O ratio.

 \begin{figure}[!b]
  \vspace{-5mm}
  \centering
  \includegraphics[width=7.5 cm]{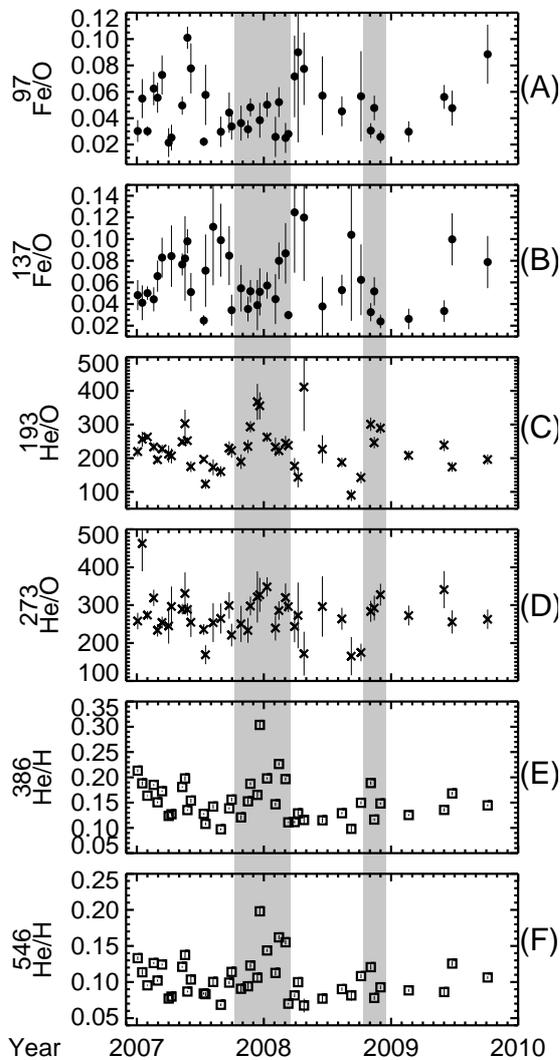}
  \caption{CIR event abundance ratios. Panels (A-B): Fe/O for 97 and 137 keV/n. Panels (C-D): He/O for 193 and 273 keV/n. Panels (E-F): He/H for 386 and 546 keV/n. {\it Shaded areas} are described in the text.}
  \label{fig02}
 \end{figure}

In Figure \ref{fig02} we explore in more detail variations of the CIR event elemental ratios in the period January 2007-December 2009. Panels (A) and (B) show Fe/O ratios for energies 97 and 137 keV/n; panels (C) and (D) He/O ratios for 193 and 273 keV/n; panels (E) and (F) He/H ratios for 386 and 546 keV/n. The wider shaded bar denotes approximate period (October 2007-February 2008) when both He/O and He/H elemental ratios show local increase (by a factor of $\sim$ 2.5) and Fe/O ratio shows decrease (by a factor of $\sim$ 5). The narrow bar marks another such period from the middle of October 2008 to the middle of December 2008. In addition to these two intervals the He/H and Fe/O ratios have a local maximum and a minimum, respectively, at the beginning of 2007. Thus, the He/H and Fe/O ratios show variations on nearly annual basis in 2007-2008. Similar, but less pronounced variations in the He/O and He/H ratios are also seen on STEREO-B. The Fe/O ratio on STEREO-B shows a random spread about the nominal value. This can be due to a poorer mass resolution in SIT-B caused by noise in the detector.

\begin{figure}[!t]
  \vspace{-5mm} 
 \centering
  \includegraphics[width=6.cm]{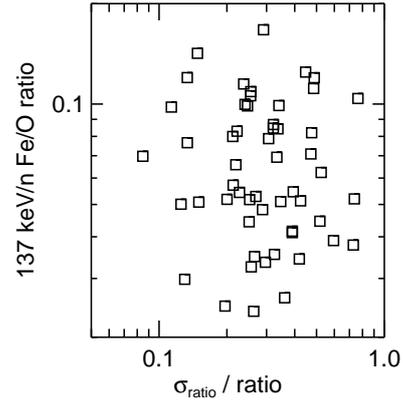}
  \caption{CIR Fe/O ratios for 137 keV/n vs. corresponding statistical errors.}
  \label{fig03}
 \end{figure}

Figure \ref{fig03} shows a scatter plot of the CIR Fe/O ratios versus corresponding standard deviations. This figure shows no relation between the ratio and its statistical error. This indicates that the trend in the variations of the Fe/O ratios seen in Figures \ref{fig01}-\ref{fig02} is not accidental. 

\begin{figure*}[t!]
  \vspace{-5mm}
  \centering
  \includegraphics[width=15. cm]{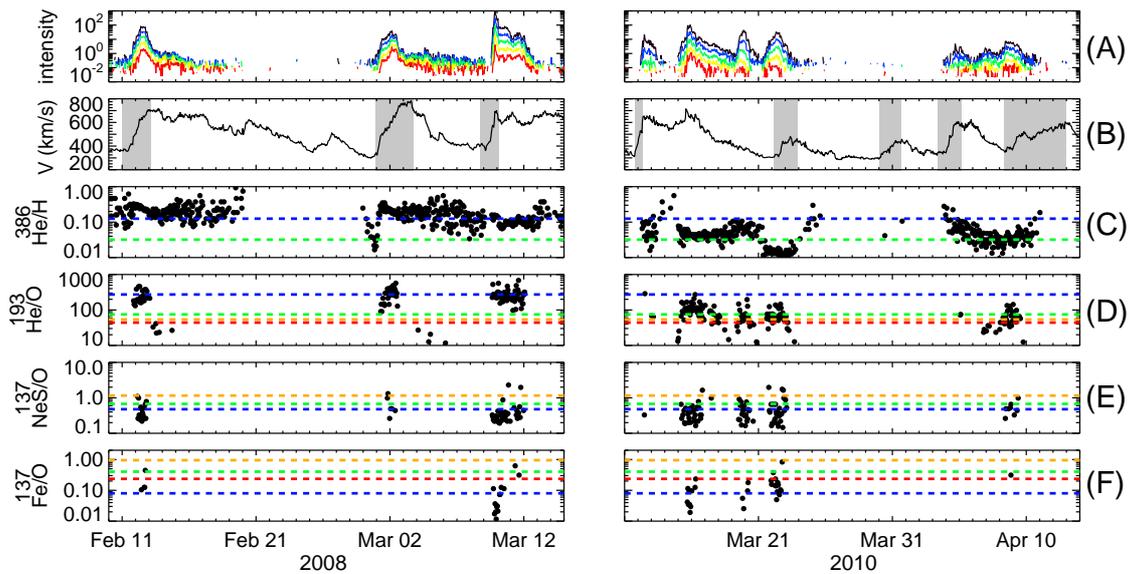}
  \caption{Panels (A): SIT/STEREO-A 1-hr averaged He intensity \#/(cm$^{2}$\,s\,sr\,MeV/n) for 189, 269, 384, 550 and 787 keV/n. Panels (B): Solar wind speed. Panels (C - F): 1-hr averaged 386 keV/n He/H, 193 keV/n He/O, 137 keV/n NeS/O and 137 keV/n Fe/O elemental ratios. {\it Grey shaded} regions mark the time intervals of the CIRs. {\it Dashed lines} present abundances in various particle populations (see Figure \ref{fig01}).}
  \label{fig04}
 \end{figure*}

Figure \ref{fig04} compares elemental abundances for the selected CIR events in 2008 and 2010. Panel (A) shows SIT-A 1-hr He ion intensity in five energy channels. Panel (B) shows solar wind speed from PLASTIC instrument \cite{lab12}. Panels (C - F) show 1-hr averages of He/H, He/O, NeS/O and Fe/O abundance ratios. The shaded bars indicate the compression regions from previously cited list of CIRs. The CIR events in February-March 2008 (left side in Figure \ref{fig04}) have characteristic corotating elemental abundances while the CIR events in March-April 2010 (right side) have He/H and He/O ratios decreased to the SEP abundances. The NeS/O and Fe/O ratios in March-April 2010 remained at corotating values and essentially do not differ from the abundances observed in February-March 2008

\section{Discussion}

Using data from the PLASTIC instrument aboard STEREO-A, Drews \etal \cite{lab13} reported on enhancements of He$^{+}$ and Ne$^{+}$ pick-up ions during the helium cone traversal around November 6, 2007 and October 1, 2008 with an approximate half width of 54 days. The authors observed that He focusing cone was more pronounced on November 6, 2007 than on the second passage around October 1, 2008.

The observations by SIT-A instrument show that the approximate start of the period of He/H and He/O enhancement and Fe/O depletion in October 2007 matches well with the start time of the pick-up ions enhancements reported in \cite{lab13}. This suggests that the enhanced He/H and He/O ratios at high energies might result in an enhanced production rate of pick-up He$^{+}$ seed population entering to the CIR acceleration. The pattern observed by SIT-A remained until January-February 2008, exceeding the period of the pick-up ions enhancement observed in \cite{lab13}. In addition to the He/H maximum at the end of 2007 we found two other maxima, one near the beginning of 2007 and other around the end of 2008. The timing of all tree increases well agree with the yearly passage of the He focusing cone.

Kallenbach \etal \cite{lab14} discussed that suprathermal He$^{+}$/He$^{2+}$ abundance ratio reflects the annual variations of the He$^{+}$ pick-up ions. In contrast, M\"{o}bius \etal \cite{lab3} and Kucharek \etal \cite{lab15} have not found signatures of the gravitational focusing cone in the observations of the He$^{+}$/He$^{2+}$ ratio in the energetic population. The authors discussed that injection and acceleration conditions masked He$^{+}$ pick-up ion variations. Note the observations reported in \cite{lab3,lab15} were acquired over the period relatively close to the sunspot maximum of the solar cycle 23. The observations reported in this survey were performed during prolonged solar minimum period under very simple solar wind conditions dominated by stably recurring CIRs \cite{lab11}. This probably led to the much more uniform injection and acceleration conditions in the CIRs making it possible to see the sign of the focusing cone.

Drews \etal \cite{lab13} noted that O$^{+}$ pick-up ions were distributed evenly in time and do not show any enhancement during focusing cone traversal. Therefore the variations of the Fe/O observed by the SIT are likely due to other causes and need further investigations. We note that the negative correlation between the He/O and Fe/O ratios, apparent from our observations, has been previously reported in \cite{lab2} with no definitive conclusion. The authors suggested temporal or solar cycle effects.

We observed a number of CIR events in 2010 with decreased He/H and He/O abundance ratios to the SEP composition while the NeS/O and Fe/O ratios remained close to the corotating abundances. It is interesting that changes in the CIR composition appeared in the period of the enhanced SEP activity. These observations are consistent with previous suggestions that CIRs reaccelerate particles from earlier SEP events \cite{lab16}. The intensity of the heavier ions in the SEP seed population was probably too low to change the NeS/O and Fe/O abundances in the reported CIR events.

\vspace{3 mm}

This work was supported by the Bundesministerium f\"ur Wirtschaft under grant 50 OC 0904. The work at the Johns Hopkins University/Applied Physics Laboratory was supported by NASA under contract SA4889-26309 from the University of the California Berkeley.


\clearpage


\begin{thebibliography}{}

\bibitem{lab1} I. G. Richardson, Space Sci. Rev., 2004, {\bf
111}: 267-376

\bibitem{lab2} G. M. Mason \etal, ApJ, 2008, {\bf 678}: 1458-1470

\bibitem{lab3} E. M\"{o}bius \etal, GRL, 2002, {\bf 29}: 1016

\bibitem{lab4} G. Gloeckler \etal, JGR, 2000, {\bf 105}: 7459-7463

\bibitem{lab5} G. M. Mason \etal, Space Sci. Rev., 2008, {\bf 136}: 257-284

\bibitem{lab6} G. M. Mason \etal, ApJ, 1997, {\bf 486}: 149-152

\bibitem{lab7} M. I. Desai \etal, ApJ, 2006, {\bf 649}: 470-489

\bibitem{lab8} J. E. Mazur \etal, ApJ, 1993, {\bf 404}: 810-817

\bibitem{lab9} G. M. Mason \etal, ApJ, 2004, {\bf 606}: 555-564

\bibitem{lab10} M. I. Desai \etal, ApJ, 2003, {\bf 588}: 1149-1162

\bibitem{lab11} R. Bu\v{c}\'{\i}k \etal, JGR, 2011, in press

\bibitem{lab12} A. B. Galvin \etal, Space Sci. Rev., 2008, {\bf 136}: 437-486

\bibitem{lab13} C. Drews \etal, JGR, 2010, {\bf 115}: 108

\bibitem{lab14} R. Kallenbach \etal, Astrophys. Space Sci., 2000, {\bf 274}: 97-114

\bibitem{lab15} H. Kucharek \etal, JGR, 2003, {\bf 108}: 8040

\bibitem{lab16} L. G. Kocharov \etal, JGR, 2003, {\bf 108}: 1404

\end{thebibliography}
\end{document}